\documentclass{emulateapj}
\usepackage{apjfonts}

\journalinfo{The Astrophysical Journal, 700:L34--L38, 2009 July 20}
\slugcomment{Received 2009 May 21; accepted 2009 June 17; published 2009
  June 30}

\shortauthors{CAMILO ET AL.}
\shorttitle{DISCOVERY OF PSR~J1747--2809 IN SNR~G0.9+0.1}

\begin{document}

\def\ergs{\,erg\,s$^{-1}$}
\def\pccc{\,pc\,cm$^{-3}$}
\def\rlum{\,mJy\,kpc$^2$}

\def\psr{PSR~J1747--2809}
\def\snr{G0.9+0.1}
\def\cxou{CXOU~J174722.8--280915}
\def\psrb{PSR~J1833--1034}
\def\snrb{G21.5--0.9}

\title{Discovery of the energetic pulsar J1747--2809 in the supernova
remnant G0.9+0.1}

\author{F.~Camilo\altaffilmark{1},
  S.~M.~Ransom\altaffilmark{2},
  B.~M.~Gaensler\altaffilmark{3},
  and D.~R.~Lorimer\altaffilmark{4}
}

\altaffiltext{1}{Columbia Astrophysics Laboratory, Columbia University,
  New York, NY~10027, USA}
\altaffiltext{2}{National Radio Astronomy Observatory, Charlottesville, 
  VA~22903, USA}
\altaffiltext{3}{Sydney Institute for Astronomy, School of Physics,
  The University of Sydney, NSW~2006, Australia}
\altaffiltext{4}{Department of Physics, West Virginia University,
  Morgantown, WV 26506, USA}

\begin{abstract}
The supernova remnant \snr\ has long been inferred to contain a central
energetic pulsar.  In observations with the NRAO Green Bank Telescope
at 2\,GHz, we have detected radio pulsations from \psr.  The pulsar has
a rotation period of 52\,ms, and a spin-down luminosity of $\dot E =
4.3\times10^{37}$\ergs, the second largest among known Galactic pulsars.
With a dispersion measure of $\mbox{DM} = 1133$\,pc\,cm$^{-3}$, \psr\ is
distant, at $\approx 13$\,kpc according to the NE2001 electron density
model, although it could be located as close as the Galactic center.
The pulse profile is greatly scatter-broadened at a frequency of 2\,GHz,
so that it is effectively undetectable at 1.4\,GHz, and is very faint,
with period-averaged flux density of $40\,\mu$Jy at 2\,GHz.

\end{abstract}


\section{Introduction} \label{sec:intro} 

With a Galactic core-collapse supernova rate of 1--3 per century
\citep[e.g.,][]{dhk+06}, young neutron stars (with age $\la 10$\,kyr)
are rare.  Nevertheless, even by those standards the known sample is
woefully incomplete.  Only 12 rotation-powered pulsars with characteristic
age $\tau_c < 10$\,kyr are known in the Galaxy, and less than 20 such
pulsar--supernova remnant (SNR) associations are firmly established.
Developing a more complete picture of the young pulsar population
contributes to an understanding of the birthrate of neutron stars and
of the physics of their creation in stellar core collapses.

While most SNRs are in principle good locations to search for young
pulsars, wholesale searches require multiple telescope pointings
each with potentially inadequate sensitivity and have had limited
success \citep[e.g.,][]{gra+96,kmj+96,llc98}.  A more recent and
successful approach has been to target, with single deep observations,
compact pulsar wind nebulae (PWNe), identified via X-ray or radio
imaging and spectroscopy, which indicate the presence of a young
neutron star even in the absence of the detection of pulsations
\citep[e.g.,][]{clb+02,cmg+02,cmgl02,crg+06,hcg+01,rhr+02}.  Detection of
the period $P$, $\dot P$, and derived quantities, underlies significant
further understanding of the pulsar, its relativistic wind, PWN, and
environment \citep[see, e.g.,][]{gs06}.

\begin{figure}[t]
\begin{center}
\includegraphics[angle=0,scale=0.42]{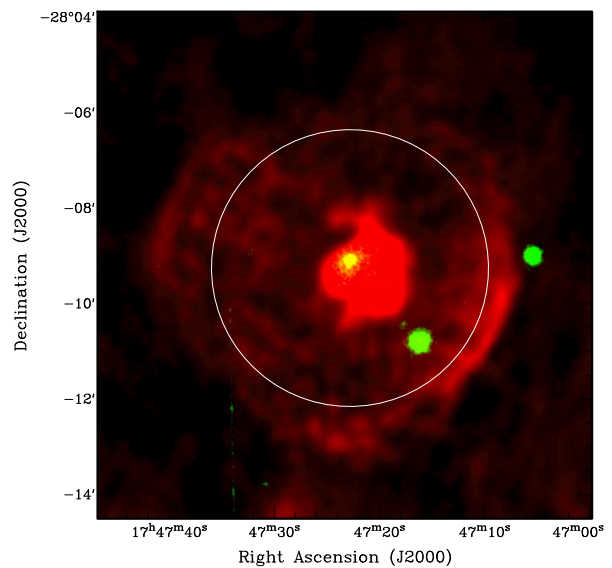}
\caption{\label{fig:snr}
A composite radio and X-ray image of SNR~\snr\ and its central pulsar
wind nebula (PWN).  In red is shown a 1.5\,GHz image, using VLA data
taken in 1984 \citep[reproduced from][]{gpg01}; in green is displayed a
0.5--10\,keV X-ray image using the EPIC PN detector on {\em XMM-Newton},
produced from an archival observation taken in 2003 \citep[see][]{sbmb04}.
Yellow denotes the radio- and X-ray-bright core of the PWN, within which
is located the hard X-ray point source \cxou\ \citep{gpg01} that likely
is emission from \psr.  The VLA data are at an angular resolution of
$15''\times11''$ while the {\em XMM-Newton}\ data have been smoothed
with a $5''$ gaussian.  The white circle shows the half-power point of
the GBT beam for the pulsar search at 2\,GHz.  The X-ray source inside
the SNR shell and to the south-west of the X-ray/radio core is the X-ray
and $\gamma$-ray transient XMMU~J174716.1--281048 = IGR~J17464--2811
\citep{dsm+07}.  The X-ray source outside and to the west of the SNR
shell is coincident with the {\em Chandra}\ source CXO~J174705.4--280859
\citep{mbbw06}.
}
\end{center}
\end{figure}

\begin{deluxetable*}{lcccccc}
\tablecaption{\label{tab:obs}Observations of \snr\ and \psr\ }
\tablecolumns{7}
\tablewidth{0.67\linewidth}
\tablehead{
\colhead{Date}                   &
\colhead{Telescope}              &
\colhead{Frequency}              &
\colhead{Bandwidth}              &
\colhead{Sample time}            &
\colhead{Integration time}       &
\colhead{$P$\,\tablenotemark{a}} \\
\colhead{}         &
\colhead{}         &
\colhead{(GHz)}    &
\colhead{(MHz)}    &
\colhead{(ms)}     &
\colhead{(hr)}     &
\colhead{(ms)}      
}
\startdata
2002 May 16 & Parkes & 1.4 & $512\times0.5$  & 2.0  & 9.3 & \nodata      \\
2002 Nov  5 & Parkes & 1.4 & $512\times0.5$  & 1.6  & 7.5 & \nodata      \\
2003 Nov  5 & Parkes & 2.9 & $192\times3.0$  & 0.25 & 5.6 & \nodata      \\
2005 Oct 20 & Parkes & 1.4 & $ 96\times3.0$  & 0.25 & 9.4 & \nodata      \\
2006 Jan  8 & GBT    & 1.9 & $768\times0.78$ & 0.08 & 6.0 & 52.137293(2) \\
2009 Mar 11 & GBT    & 2.0 & $512\times1.56$ & 0.16 & 5.8 & 52.152855(2)
\enddata
\tablenotetext{a}{The uncertainties in barycentric periods, given on the
last digit in parentheses, are the $1\,\sigma$ values obtained from
TEMPO fits. }
\end{deluxetable*}

The composite SNR~\snr\ consists of a radio shell $8'$
in diameter surrounding a $2'$ PWN (Helfand \& Becker
1987; see Figure~\ref{fig:snr})\nocite{hb87}.  Based on
the very high interstellar absorption \citep[$N_H \approx
1.3\times10^{23}$\,cm$^{-2}$;][]{gpg01,pdw03,smib00,sbmb04}, its
distance is large, here parameterized by $d_{10} = d/\mbox{(10\,kpc)}$.
The PWN is luminous in radio, with $L_r\,(10^7-10^{12.4}\,{\rm Hz})
= 1.7\times10^{35}d_{10}^2$\ergs\ \citep{dgd08}, and filled with
X-ray synchrotron emission with $L_X\,(2-10\,{\rm keV}) = 0.4\,L_r$
\citep{pdw03}.  It is also a very-high-energy $\gamma$-ray source,
with $L_{\gamma}\,(>0.2\,{\rm TeV}) = 0.4\,L_X$ \citep{aaa+05a}.
Based on these energetics and a variety of empirical relations,
it has been predicted that the pulsar powering this PWN has $P \sim
0.1$--0.2\,s and spin-down luminosity $\dot E \sim 2\times10^{37}$\ergs\
\citep{dgd08,mfg+09,smib00}, while the SNR shell size implies an age
of about 1--7\,kyr \citep{msi98}.  The hard X-ray point source \cxou,
with 1\% the luminosity of the PWN and surrounded by small-scale ordered
structure, is likely emission from the pulsar \citep{gpg01}.  We have
searched this location for a radio pulsar, and in this Letter report
the discovery of \psr, the central pulsar in SNR~\snr.

\section{Observations and Results} \label{sec:obs} 

We began our search for the pulsar in SNR~\snr\ at the ATNF Parkes
telescope, where during 2002--2005 we used three combinations of search
frequencies and filterbank data acquisition systems to do four very
long observations (see Table~\ref{tab:obs}).  The \citet{cl02} NE2001
free electron distribution model predicts that in this direction, for
$d_{10} = 1$, the expected dispersion measure would be $\mbox{DM} =
750$\,pc\,cm$^{-3}$, and that the pulse broadening due to interstellar
scattering at the standard search frequency of 1.4\,GHz would be
$\tau_{1.4} \ga 10$\,ms.  Given the possibility of larger actual DM and
$\tau_{1.4}$, one of our observations was at 3\,GHz.

We analyzed all search data with standard pulsar search techniques
implemented in PRESTO \citep{ran01,rem02}, including the excision of
radio frequency interference (RFI) and a nearly optimal set of trial
DMs \citep[for more details see, e.g.,][]{crg+06}.  We dedispersed at
up to twice the maximum Galactic DM of 1580\,pc\,cm$^{-3}$ predicted by
NE2001 in this direction.  No new pulsars were detected in any of the
Parkes data sets.

The Parkes sensitivity limits at 1.4\,GHz correspond to luminosity
$L_{1.4} \equiv S_{1.4}\,d^2 \la 7\,d_{10}^2$\,mJy\,kpc$^2$, provided
that scattering was not the limiting factor.  At 3\,GHz, although the sky
background temperature was reduced, any ordinary pulsar would have had
an even more significantly reduced flux, and the equivalent luminosity
limit was worse.  Since young pulsars can have luminosities at least as
small as $L_{1.4} \approx 0.5$\,mJy\,kpc$^2$ \citep{csl+02}, we did a
deeper search at the NRAO Green Bank Telescope (GBT).

In 2006 January we observed \snr\ at the GBT at a central frequency of
1.95\,GHz using the Spigot autocorrelation spectrometer \citep{kel+05}.
For a typical spectral index of $\approx -1.6$ \citep{lylg95}, the
pulsar would be fainter by a factor of about 1.7 by comparison with
1.4\,GHz.  However, this was more than made up by the larger gain and
bandwidth at the GBT, such that our observation was a factor of about
2 more sensitive than the best Parkes search, both compared at the same
fiducial frequency of 1.4\,GHz.  In addition, by using a higher search
frequency, the scattering timescale was reduced by a factor of about 4,
which proved crucial in light of the actual pulsar period and the amount
of observed scattering.

\begin{figure}[t]
\begin{center}
\includegraphics[angle=0,scale=0.95]{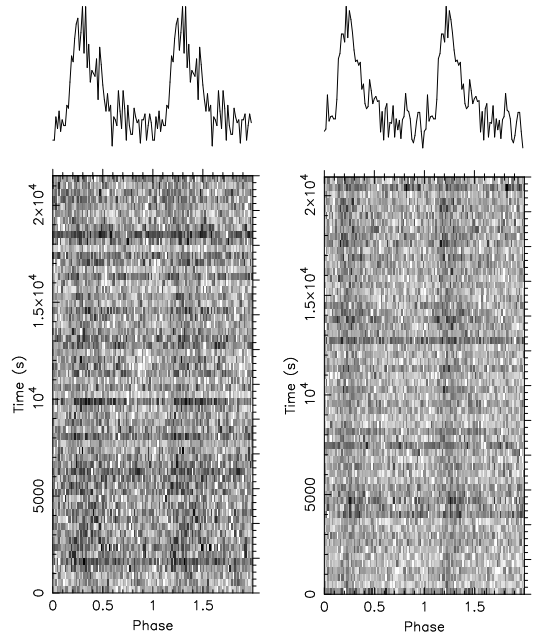}
\caption{\label{fig:psr}
Detections of \psr\ at GBT.  Each 2\,GHz profile ($P=52$\,ms) is repeated
in phase, and is shown as a function of time ({\em bottom\/}) and as
the sum of more than 400,000 pulses ({\em top\/}).  Left: discovery
observation from 2006 January 8 with Spigot.  Right: confirmation
observation from 2009 March 11 with GUPPI.  The observed pulse FWHM
increases from $0.14\,P$ at 2.35\,GHz to $0.30\,P$ at 1.75\,GHz, due to
interstellar scattering.
}
\end{center}
\end{figure}

We identified in these data a highly dispersed ($\mbox{DM} =
1145$\,pc\,cm$^{-3}$) and scattered pulsar candidate with period
$P=52$\,ms (Figure~\ref{fig:psr}).  It is a faint signal, with flux
density $S_{2} = 45\,\mu$Jy, compared to the detection threshold
in this observation for a pulsar with these characteristics of
approximately $20\,\mu$Jy.  We confirmed the pulsar in an equivalent
observation using the Green Bank Ultimate Pulsar Processing Instrument
(GUPPI)\footnote{https://wikio.nrao.edu/bin/view/CICADA/GUPPiUsersGuide.}
on 2009 March 11 (Figure~\ref{fig:psr}), for which $S_{2} = 35\,\mu$Jy.
In the presence of RFI, GUPPI, an FPGA-based digital spectrometer, has
vastly improved performance compared to Spigot.  This proved vital, since
the RFI environment has degraded enormously in the intervening 3 years.

To measure the amount of scattering that clearly affects the pulse
profile of \psr\ (Figure~\ref{fig:psr}), we did a fit to the GUPPI
profile simultaneously in seven 100\,MHz-wide sub-bands, assuming that
the scattering timescale $\tau_{\nu}$ scales with observing frequency
as $\nu^{-\alpha}$ with $\alpha = 4$ \citep[see, e.g.,][]{bcc+04}.
We obtain $\tau_1 = (0.21\pm0.03)$\,s, scaled to the usually reported
frequency of 1\,GHz (for $\alpha$ in the range 3.6--4.4, $\tau_1$ varies
over 0.16--0.28\,s).  This compares to $\tau_1 = 0.12$\,s predicted by
\citet{cl02} for this DM and direction.  The observed DM is biased by
the large degree of scattering, and in our fitting process we obtain
the corrected $\mbox{DM} = (1133\pm3)$\pccc.

A comparison of the barycentric periods for the discovery and confirmation
observations (Table~\ref{tab:obs}) shows that the pulsar period increased
by $\Delta P = (15.562\pm0.003)\,\mu$s.  The time interval between these
observations was $\Delta T = 1157.8$ days (MJD $54901.5 - 53743.7$).
The average period derivative is thus $\dot P = \Delta P/\Delta T =
(1.5557\pm0.0003)\times10^{-13}$.  The corresponding derived pulsar
parameters are $\dot E = 4.3\times10^{37}$\ergs, $\tau_c = P/(2\dot
P) = 5.3$\,kyr, and surface magnetic dipole field strength $B =
2.9\times10^{12}$\,G.  For the observed DM, the NE2001 model yields $d
\approx 13$\,kpc, though with substantial uncertainty.

\psr\ has a magnetic field strength at its light cylinder of $B_{\rm
lc} = 1.9\times10^5$\,G, which is the seventh highest among pulsars
known in the Galactic disk.  It has been suggested that $B_{\rm lc}$
is a controlling parameter for the emission of very narrow ``giant''
radio pulses \citep{cstt96}, and among the four pulsars known to emit
giant pulses, PSR~B0540--69, in the LMC, has the smallest value, $B_{\rm
lc} = 3.6\times10^5$\,G.  Therefore, in addition to the periodicity
searches of the data listed in Table~\ref{tab:obs} (we also did a
more sensitive folding analysis of all the Parkes data sets, using the
subsequently determined values of $P$ and $\dot P$, but did not detect
the pulsar), we carried out a search for dispersed individual pulses
in the two GBT data sets.  We used standard matched-filter techniques
described in detail by \citet{cm03} and implemented in the SIGPROC pulsar
search package\footnote{http://sigproc.sourceforge.net.}.  No single
pulses were detected.  However, our sensitivity was much degraded as a
consequence of the large single-channel dispersion smearing (1--2\,ms)
and the huge scattering (about 13\,ms at 2\,GHz), but for which we could
easily have detected giant pulses like those observed from PSR~B0540--69
by \citet{jr03}.

\section{Discussion} \label{sec:disc} 

\psr\ is a rare young neutron star in that its spin-down luminosity
is surpassed among known Galactic pulsars only by the Crab.  We have
estimated the chance probability of detecting such a pulsar in our
search by simulating an inner-Galactic GBT survey with individual 6\,hr
Spigot observations at 2\,GHz.  Considering the best population model
of \citet{lfl+06}, we find that such a survey would detect one ordinary
(non-millisecond) pulsar every 0.07\,deg$^2$.  Our beam area is 11\% of
this, which is also the probability of finding one such pulsar blindly.
Until now, only the Crab was known with $\dot E > 4\times10^{37}$\ergs\
among 1600 ordinary Galactic pulsars.  The probability of finding any such
high $\dot E$ pulsar by chance is thus $\sim 0.11/1600 = 7\times10^{-5}$.
\psr\ is also very distant, based on its large DM, as expected for
the pulsar in SNR~\snr\ from the high $N_H$ to its PWN.  The current
positional uncertainty of \psr\ is $\pm3'$, which is sufficient to
locate it in projection within the \snr\ shell (Figure~\ref{fig:snr}).
Within this area, there are no radio or X-ray sources other than the
\snr\ PWN that could plausibly account for a young pulsar with $\dot E
= 4\times10^{37}$\ergs, despite the great sensitivity of existing VLA,
{\em Chandra X-ray Observatory\/}, and {\em XMM-Newton\/} observations.
There is therefore no doubt that \psr\ and \snr\ are associated.

The electron density model is not well constrained toward the distant
inner Galactic regions \citep[see][]{cl02}.  Prior to the discovery of
\psr, only nine pulsars were known with $\mbox{DM}>1100$\pccc, of which
only two are within 20 degrees of the Galactic center (GC), none with
independently estimated distances \citep{mhth05}.  Therefore, the NE2001
distance of 13\,kpc for \psr\ could be substantially in error.  It is not
excluded that the pulsar could be located at 8--9\,kpc, physically near
the GC.  However, given both the very large $N_H$ and DM, it is unlikely
that the pulsar is located substantially closer than the GC.  Also, the
ratio $N_H/\mbox{DM} \approx 40$ for \psr\ is higher than seen toward all
but about four other pulsars \citep[see][]{gvc+04,crg+06}.  Gaensler et
al.\ argue that this indicates a location behind substantial intervening
molecular material, which would not be surprising for a location near or
beyond the GC.  It is also intriguing that two pulsars located only 0.3
degrees in projection from the GC (at a lateral distance of about 40\,pc
if near it) have $\mbox{DM} \approx 1100$\pccc\ \citep{jkl+06}, like \psr,
which among known pulsars is the third nearest to the GC in projection.
The scattering timescale for those two pulsars is a factor of a few larger
than for \psr, but still very small compared to the levels expected if
they were located within the scattering screen thought to surround the GC
at a distance estimated as 50--330\,pc by \citet{lc98}.  \citet{jkl+06}
therefore argue that those two pulsars are located somewhat in front of
the GC.  The same may apply to \psr, with a larger lateral separation of
about 130\,pc, but we also cannot exclude a substantially larger distance.
We consider it likely that $0.8 \la d_{10} \la 1.6$.

The spin parameters of the new pulsar are similar to those of \psrb\
in SNR~\snrb, which has $P=61$\,ms, $\dot E = 3.4\times10^{37}$\ergs,
$\tau_c = 4.9$\,kyr, and $d = 4.7$\,kpc \citep{crg+06,tl08}.  It is
therefore of interest to compare the properties of both systems.
The radius of the central PWN in \snr\ is $R_{\rm PWN} = 3\,d_{10}$\,pc,
nominally 50\% larger than the PWN in \snrb.  The radio luminosities
may be similar, although this comparison is uncertain because of the
much better frequency coverage for \snrb\ \citep[see][and references
therein]{dgd08,bwd01}.  In X-rays (2--10\,keV), $L_X \mbox{(G0.9+0.1)}
\approx 0.7\,d_{10}^2\,L_X \mbox{(G21.5-0.9)}$ \citep[][]{pdw03,scs+00}.
At the highest energies, both PWNe are unresolved TeV sources, and
colocated with their lower-energy counterparts, but \snr\ is $\approx
5\,d_{10}^2$ times more luminous than \snrb\ in a comparable band
\citep{aaa+05a,ddt+08,gcd+08,mfg+09}.

Although the characteristic ages of both PSRs~J1747--2809 and J1833--1034
are $\tau_c = 5$\,kyr, the actual age of \psrb\ and its SNR is only
$\tau \approx 1$\,kyr \citep{bb08,bvcb05,crg+06}.  \psr\ could therefore
be the older of the two, and one might be tempted to appeal to this
possible difference to explain the larger size of the \snr\ PWN (and
of its shell, whose radius $R_{\rm SNR}$ is also nominally 50\% larger
than that of \snrb) and (along with a smaller nebular magnetic field) its
possibly lower X-ray efficiency.  However, the nominal difference in the
two efficiencies is small compared to the scatter observed among young
pulsars.  Dissimilar SN explosion kinetic energies and ejected masses
could instead explain the different radii \cite[see, e.g.,][]{vagt01}.
Some observables point to the possibility that also for \snr, $\tau <
\tau_c$.  The \snr\ shell is relatively bright in radio, unlike \snrb.
This could be due to a different circumstellar environment, or may
suggest that the SNR has swept enough mass and is transitioning to the
adiabatic phase.  Nevertheless, the PWN still retains approximate circular
symmetry and central location within the SNR shell, which likely indicates
that a strong reverse shock has not yet formed.  In that case, for the
observed ratio $R_{\rm PWN}/R_{\rm SNR} \approx 0.25$ (similar to that of
\snrb), the PWN evolutionary models of \citet{bcf01} indicate a small age.
The PWN energetics also indicate a small age: for \snr, the PWN magnetic
and particle energies add to $\sim 10^{48}\,d_{10}^2$\,erg \citep{dgd08}.
This is smaller than $\dot E \tau_c = 7\times10^{48}$\,erg, suggesting
that $\tau \ll \tau_c$ \citep[see][]{che05}.  All these ideas are outlined
more fully for \psrb/\snrb\ by \citet{crg+06}, including a discussion of
the very significant differences with PSR~J0205+6449 and its PWN 3C~58,
despite comparable spin parameters.  All these considerations point in
the case of \psr/\snr\ to a system that, while not necessarily quite as
young as \psrb/\snrb, may have an age of no more than about 2--3\,kyr.
In turn, for spin evolution under constant magnetic moment with braking
index in the observed range 2--3 \citep{lkg+07}, this would imply a
birth period of $\ga 40$\,ms.

The TeV $\gamma$-ray emission observed from both \snr\ and \snrb\ is
most likely due to inverse Compton scattering of relativistic pulsar
wind electrons.  The seed photons for such scattering in general arise
from dust, the cosmic microwave background, and star light.  If \psr\
is located close to the GC, the main contribution to the photon field
in \snr\ likely originates in star light \citep{aaa+05a}, and this could
account for much of the greater TeV luminosity of \snr\ compared to \snrb.
If on the other hand \psr\ is located at a substantially greater distance,
we would infer an even larger TeV luminosity for \snr\ while the stellar
photon field would have lower energy density.  Thus, careful modeling of
the TeV emission from the \snr\ PWN may help to constrain the distance to
the system.  In any case, it may be that much of the observed difference
between \snr\ and \snrb\ arises not so much from intrinsic differences
in the particle spectrum injected, respectively, by PSRs~J1747--2809
and J1833--1034, as from their different environments.

The pulsars themselves have comparable luminosities and are very
faint.  For \psrb, $L_{1.4} \approx 2$\,mJy\,kpc$^2$ and $L_X \approx
3\times10^{-5} \dot E$ \citep{crg+06}.  For \psr, $L_{1.4} \approx
7\,d_{10}^2$\,mJy\,kpc$^2$, and $L_X \approx 1.5\times10^{-5}\,d_{10}^2
\dot E$, assuming that \cxou\ is the counterpart \citep[another potential
candidate lies $10''$ to its north;][]{gpg01}, which may be tested
with the determination of a pulsar position via timing measurements.
At GeV energies, pulsations have already been detected from \psrb\
with the Large Area Telescope on the {\em Fermi Gamma-ray Space
Telescope\/}\footnote{http://moriond.in2p3.fr/J09/transparents/caliandro.ppt.}.
\psr\ is, at least, at twice the distance, and near the bright GC,
and no detection by {\em Fermi\/} has yet been reported.

Because of interstellar scattering, \psr\ is not detectable at 1.4\,GHz
without a much more sensitive observation than is possible in practice
($\tau_{1.4} \approx 55$\,ms, estimated by scaling from $\tau_1$).
Until recently, there were only four Galactic young pulsars known with
$P<61$\,ms (the Crab, J1913+1011, B1951+32 in SNR~CTB~80, and J2229+6114
associated with 3EG~J2227+6122).  The bright PSR~J1410--6132, with
$P=50$\,ms and $\mbox{DM} = 960$\,pc\,cm$^{-3}$, but very scattered
at 1.4\,GHz and discovered instead at 6\,GHz \citep{ojk+08}, and the
similarly scattered but $\sim 100$ times fainter \psr, show the importance
of doing the utmost in dedicated searches of interesting objects to
plumb the depths of the pulsar luminosity distribution and to minimize
other biases against detecting short period neutron stars \citep[see
also][for detection of other highly scattered pulsars]{crhr07,jkl+06}.
\psr\ lies within the group of low luminosity young pulsars ($L_{1.4}
\la 10$\,mJy\,kpc$^2$), nearly all of which were discovered in deep
directed searches.  These now make up one-third of all known young pulsars
\citep[see discussion in][]{crg+06}, and there is little reason to suppose
that such pulsars are intrinsically rare.  The future, if not bright,
appears promising.

\acknowledgements
The GBT is operated by the National Radio Astronomy Observatory, a
facility of the National Science Foundation operated under cooperative
agreement by Associated Universities, Inc.  We thank the staff at NRAO
for quickly approving and scheduling the Rapid Response Science request
to confirm the pulsar.  The Parkes Observatory is part of the Australia
Telescope, which is funded by the Commonwealth of Australia for operation
as a National Facility managed by CSIRO.  This work was supported in part
by the NSF through grant AST-0908386.  B.M.G.\ acknowledges the support
of a Federation Fellowship from the Australian Research Council through
grant FF0561298.  D.R.L.\ is partially supported by West Virginia EPSCoR
through a Research Challenge Grant.

{\em Facilities:} \facility{GBT (GUPPI, Spigot)}, \facility{Parkes
(PMDAQ)}

\end{document}